# Non-resonant radio-frequency response in superconducting $MgB_2$

S. Sarangi[*] and S. V. Bhat

Department of Physics, Indian Institute of Science, Bangalore-560012, India

[*]Corresponding author:

Subhasis Sarangi

Department of Physics

Indian Institute of Science

Bangalore – 560012, India

Tel.: +91-80-22932727, Fax: +91-80-23602602

E-mail: subhasis@physics.iisc.ernet.in

**Abstract:**

Non-resonant radio frequency absorption (NRRA) of superconducting $MgB_2$ pellet is investigated as a function of frequency, temperature, pressure, rf power, magnetic field, sample size and grain size. The NRRA signals of granular $MgB_2$ samples show three separate and distinguish temperature dependent phase reversals, one of them is similar to the phase reversal occurs very often in high $T_c$ superconductors at temperature few degrees below $T_c$, the other two are anomalous and occur at lower temperatures. Interpretation based on the assumptions in the line shape model indicates that the Josephson junctions are weakly coupled in $MgB_2$ than other high temperature superconductors like $YBa_2Cu_3O_7$, $La_{1.9}Sr_{0.1}CuO_4$ and $Bi_2Sr_2CaCu_2O_8$. The samples show a significant magnetic field sweep hysteresis, which is a signature of trapped flux. The hysteresis behavior suggests that the vortices in the $MgB_2$ are rather rigid. The effects of passing dc current through the sample on the NRRA signal patterns are discussed. It is seen that the transition temperature $T_c$ measured using the NRRA technique decreases with decreasing sample size, whereas the $T_c$ measured using ac susceptibility and temperature dependent resistivity is independent of the sample size. The results are discussed in terms of resistively shunted junction, flux flow and bond percolation models. Effect of MgO doping is analyzed and discussed. The intergranular coupling energy and the activation energy of $MgB_2$ are calculated and compared with other high temperature superconductors. The effects of sintering on the weak links in $MgB_2$ polycrystalline pellet are discussed.

**Introduction:**

Non-resonant radio frequency absorption (NRRA) is a highly sensitive and non-invasive technique like NRMA (non-resonant microwave absorption) to detect and characterize the superconducting phase [1, 2, 3, 4, 5, 6, 7, 8]. Superconductors exhibit energy loss when exposed to time-varying magnetic fields or transport current like rf and microwave, the so-called ac loss in superconductors. The rf response investigation is one of the effective method in the investigation of flux dynamics because it is highly sensitive and can give information on the inter-grain and intra-grain pinning features. It also gives the detail information regarding the weak link responses. The motion of the vortices over pinning centers (flux creep) and the Josephson junction decoupling in the superconductor induce dissipation. Flux creep is greatly affected by the presence of electromagnetic waves like rf and microwave.

We have carried out NRRA studies on superconducting $MgB_2$ and a sample containing ~10% by weight of MgO in $MgB_2$. The fundamental superconducting parameters of $MgB_2$ such as the upper critical field $B_{c2}$ = 12.0-18 T, the lower critical field $B_{c1} \approx 200$ G, the Ginzberg-Landau parameter $\kappa \approx 26$, the coherence length $\xi(0)$ = 4-5 nm, the penetration depth $\lambda$ = 140-180 nm, have been obtained. The compound is a very good metal in the normal state [9] and has large activation energy [10]. The compound is free from weak links [11] and is associated with fast flux creep [12]. The compound has a long coherence length in the superconducting state. These properties make it a promising candidate for applications such as current-carrying wires, tunnel junctions, microwave and rf devices.

In case of high $T_c$ superconductors (HTSC), two processes mainly contribute to the field dependent dissipations. One, the decoupling of the Josephson junctions (JJs) due to the magnetic field [13] and the other, the Lorentz force driven motion of the quantized flux lines [14], the so-called fluxons. In type II superconductors, flux motion is driven by the Lorentz force, which is determined by applied current, magnetic field and the angle between them. The decoupling of the Josephson junctions depends on $I_c$, the critical current and the JJ decoupling energy $E_j$, the energy required to break a Josephson junction. So the applied rf current can influence the transport properties in the presence of magnetic field. Direct evidence of flux motion and Josephson junction decoupling and the influence of the flux motion and resistively shunted junction on the properties of $MgB_2$ have not been addressed adequately in the literature. A detailed study of the temperature dependence of the phase, amplitude and peak-to-peak width of the NRRA signals has been carried out. The rf absorption line shape, the phase reversals and the strong hysteresis behaviors are discussed in terms of the models for the rf loss in intergranular Josephson junctions involving boundary and rf currents and flux motion.

It is reported recently that the activation energy $U_c$ in the magnesium diboride is proportional to the thickness of sample and saturates at collective pinning length $L_c$ [10]. It has been reported that the critical current density derived from magnetic measurement in the magnesium diboride superconductors depends on the sample size [15]. We studied the NRRA response of $MgB_2$ samples with different crystalline properties, sample sizes, grain sizes and pressures and sintered under different physical and chemical conditions, which are important for commercial applications. The effects of frequency and amplitude of the ac current on the NRRA signals are discussed in detail. Three separate and distinguish temperature dependent phase reversals are observed in the NRRA signals of $MgB_2$, one of them is similar to the phase reversal occurs very often in the high $T_c$ superconductors at temperature few degrees below $T_c$, the other two are anomalous and occur at lower temperatures. The results are discussed in terms of Josephson junction decoupling (JJD), resistively shunted junction (RSJ), flux flow (FF) and bond percolation (BP) models.

**Experiment:**

The starting material used in this study was the powder of $MgB_2$ (Alpha Aesar of John Matthey GmBH, Germany), which is commercially available. Our polycrystalline samples were synthesized from those powders under high pressure at high temperature as described in Ref. [16]. The material was found to be single phasic as determined by x-ray diffraction with almost theoretical density of 2.63 g/cm$^3$. All the samples show sharp superconducting transition temperature at ~ 39 K and transition width is less than 2 K as determined by $\rho \sim T$ and ac susceptibility measurements. In order to investigate the effects of different particle sizes, a graded set of samples was prepared using a differential sedimentation method as described in [17]. The samples produced by this method had a graded range of particle sizes up to known maxima: less than 30 μm, less than 25 μm, less than 20 μm, less than 15 μm and less than 10 μm. We study the effect of pressure on the NRRA signals in the granular $MgB_2$ sample by making pellets from the powder sample (grain size less than 30 μm) at three different pressures: 10 MPa, 20 MPa and 30

MPa. To study the effect of sintering on the NRRA signal, we prepared two pellets made of polycrystalline samples of $MgB_2$. Both the pellets were prepared from the same phase of $MgB_2$ powder (grain size less than 30 μm) with a pressure of 10 MPa. One of the two pellets was sintered at 650 $^0C$ in flowing Argon for 10 hours to minimize the presence of Josephson junctions. The second pellet was not sintered to enhance the possibilities of Josephson junctions in it. The sizes of the pellets (cylindrical pellets of length: 15 mm and radius 3 mm) were kept constant for all the measurements except sample size dependent experiments. The applied rf frequency and the rf amplitude were set at 10 MHz and 1 V for most of the measurements.

The MgO mixed $MgB_2$ sample was prepared as follows: stoichiometric mixture of Mg powder (99.8%) and crystalline boron 325 mesh, were ground and palletized and placed inside Ta crucible and sealed in Ar atmosphere. This was further sealed in a quartz tube with 350 mbar Ar pressure and heat treated at $950^0$ C in a box furnace for 2 h and air quenched. The sample had about 10% of MgO, which came from crystalline boron containing oxygen impurity. The transition temperature of MgO mixed $MgB_2$ is 35 K and transition width is less than 2 K as determined by $\rho \sim T$ and ac susceptibility measurements.

We have used the technique of non-resonant electromagnetic absorption to study the magnetic field dependent rf dissipation. It involves subjecting the sample to rf in a preferred orientation and scanning the magnetic field from negative to positive value through zero at a fixed temperature below $T_c$ [1, 2]. It is based on the observation that the magnetic field dependent losses in the superconducting state of a sample give rise to an intense signal when studied using conventional continuous-wave (CW) nuclear magnetic resonance (NMR) spectroscopy equipped with a low rf level Robinson oscillator operating in the 5-30 MHz range [1, 2]. The rf magnetic field was fixed at $H_{rf} \sim 10$ mOe peak to peak and oriented perpendicular to the dc magnetic field $H$ ( $\parallel c$). Magnetic field modulation and lock-in detection customarily used in CW NMR spectrometers result in the derivative $dP/dH$ of the dissipated rf power $P(H)$ recorded as a function of the field. The modulation frequency was 100 Hz and the amplitude was set at 10 G for most of the experiments. Changing the driving current into the Helmholtz coils inside the main magnet can vary the modulation amplitude. The modulation amplitude was varied 0-20 G to study the effect of modulation on the NRRA signal patterns. The incident rf power to the sample is proportional to the rf amplitude of the rf oscillations produced across the rf coil. So it is possible to pump more rf power to the sample inside the rf coil by changing the amplitude of rf oscillation. All the NRRA measurements were performed with the applied field parallel to the longest direction of the sample (a axis). An Oxford Instrument CF200 continuous He flow cryostat was used to change the temperature. The schematic of the sample region is illustrated in Figure 1. The sample and the rf probe were placed inside an Oxford instrument cryostat. The sample is kept inside the rf coil. When the sample is kept inside an rf coil, the rf magnetic field generated inside the coil induces rf current in the sample. This rf current flows through the sample. Any loss due to the passing of rf current in the sample is reflected in the NRRA signal. The samples under investigation are placed in the core of the coil forming the inductance $L$. The NRRA measurements were performed on both the sintered and non-sintered pellet each having the same size. The NRRA was measured in the magnetic field range of -150 G < $H$ < 150 G for all the measurements. Measurements were made after stabilizing the temperature for about 10 min prior to each reading in the temperature range 10 – 50 K. We study the effect of superimposing a dc current with the applied rf current on the NRRA signals. The dc current was passed across the sample by making two probe contacts at the opposite ends of the sample. The current direction was same as the direction of rf field inside the coil. We use a Keithely current source (Model 220) for the variation of the dc current. The ac susceptibility was measured with a standard three-coil pickup system and lock-in technique. The ac field strength and frequency were typically 5 Oe and 90 Hz respectively for the ac susceptibility measurements. Resistivity measurements were made in a standard four-probe geometry using Epotek H20E silver epoxy to make contacts. The contact resistance was approximately 5 Ω. Given the well-defined geometry of the samples, accurate measurements of resistivity were possible.

**Results & Discussion:**

The temperature at which (d$P$/d$H$) versus $H$ disappears is treated as the critical temperature of the sample. The anomalous appearance of the three phase reversals and the strong hysteresis behavior (The area between the forward and reverse sweeps of the magnetic field is a measure of hysteresis.) in MgB$_2$, and the weak Josephson junction signals, which observed in all the samples irrespective of their sizes, are discussed. The rf losses are independent of magnetic field in the normal state but become field dependent in the superconducting state. In order to explain the experimental results presented in this section, one must compare different mechanisms, which can give rise to dissipation in magnetic field for high-$T_c$ superconductors. The main models for the dissipations at rf and microwave frequencies given in the literature are connected to Josephson junction decoupling, flux flow, flux creep, phase slippage, and dephasing of the Josephson junctions.

NRRA signal patterns in MgB$_2$

In Figure 2, *dP/dH* versus *H* curves are shown for an MgB$_2$ pellet (cylindrical pellet of length: 15 mm and radius 3 mm, pressurized at 30 MPa, grain size is less than 10 μm) at various temperatures below $T_c$ (~39 K, determined from our field dependent rf absorption measurements) during heating. The magnetic field is scanned both in the forward and reverse directions from -150 to 150 Gauss. It is observed that the magnetic field dependent NRRA signals of MgB$_2$ sample do not show flux flow signatures at temperatures below $T_c$. Flux flow signals are always seen in other high $T_c$ materials like YBa$_2$Cu$_3$O$_7$ (YBCO), La$_{1.9}$Sr$_{0.1}$CuO$_4$ (LSCO) and BSCCO (Bi$_2$Sr$_2$CaCu$_2$O$_8$) (see Figure 3 for the presence of flux flow signals in YBCO polycrystalline sample) below the critical temperature. Flux flow signals have their own distinct features, which can be easily identified from the temperature dependent NRRA signal patterns. These signals appear only above $H_{c1}$ (The lower critical field) and those are relatively broader than JJ signals at lower temperatures. It is important for us to note that the signal patterns in Figure 2 show that the NRRA responses in MgB$_2$ are only due to the JJ decoupling and not due to the flux motion. Figure 3 shows the NRRA signals of an YBCO polycrystalline sample (cylindrical pellet of length: 15 mm and radius 3 mm) at some selected temperatures below $T_c$ (89 K). The signals are due to both flux motion and JJ decoupling. In this figure, the flux flow signals can be easily observed by looking at the movements of the peak (indicated by the arrows). The peak slowly moves towards the center of the signal with increasing temperature. This is due to the decreasing of $H_{c1}$ with increasing temperature. The amplitude of the NRRA signals of YBCO is larger than $MgB_2$ of same size throughout the temperatures below $T_c$. It is observed that not only the absolute value of dissipation is more in other HTSC like YBCO, LSCO, BSCCO than in $MgB_2$ of same size but also the change in the magnetic field dependent dissipation due to flux flow is more and easily determined by our NRRA technique throughout the temperature range in HTSC.

It is observed recently that the flux flow losses disappear if the sample length is less than the collective pinning length $L_c$ [10, 15]. In HTSC, vortices are pinned collectively by an array of point pinning centers. A longer sample accommodates longer vortices and these vortices are pinned by more pinning centers because the vortices are rigid, the pinning potential for individual vortices increases linearly with the sample length below $L_c$. For the samples longer than $L_c$, vortex start breaking up into segments [18], which are pinned individually and increase of their length will no longer result in the increase of the pinning potential. It has been shown in recent reports [19, 20] that the pinning potential increases with the sample length below the collective pinning length in the direction parallel to the field $L_c \approx (\varepsilon_0 \xi)^{2/3} / \gamma$, with $\varepsilon_0$ the basic energy scale, $\xi$ the coherence length and $\gamma$ the parameter of disorder strength, respectively. The collective pinning length of $MgB_2$ is calculated to be $\cong 1$ mm [10]. So the sample bigger than $L_c$ will have same pinning potential and activation energy irrespective of their sizes. To eliminate the possibilities and the effect of pinning length on sample size we have done all the NRRA experiments on samples with size more than the collective pinning length ($\cong$ 1 mm for MgB$_2$ sample).

It is important to understand the weak Josephson junction decoupling signals and the absence of flux flow signals in MgB$_2$. The activation energy $U_c$ for MgB$_2$ sample bigger than 1 mm ($L_c$) is about 10 times greater than the average activation energy in high temperature superconductors like YBCO, LSCO and BSCCO of similar size [10]. This high activation energy make MgB$_2$ vortex stronger and more rigid comparative to other HTSC. Due to this rigid vortex, $\eta$ the fluxon viscosity in MgB$_2$ medium increases. In the high-temperature cuprate superconductors, such as YBCO, LSCO and BSCCO, the activation energy of vortices are very low which is determined by their elastic property. Due to this low activation energy of vortices, $\eta$ the fluxon viscosity becomes very low and due to this lower $\eta$ value, the dissipation due to flux motion is higher in superconductors like YBCO, LSCO and BSCCO than MgB$_2$. To get a quantitative relation between fluxon viscosity and the energy dissipation we follow flux flow model for high frequency dissipation where the real part of the surface impedance that leads to dissipation is

$$R_s = X_0 \left[ \left( -1 + \left( \sqrt{(1 + f^2 B^2 / B_0^2)} \right) \right) / 2 \right]^{1/2}$$

and the imaginary part resulting in the change in frequency is

$$X_s = X_0 \left[ \left( 1 + \left( \sqrt{(1 + f^2 B^2 / B_0^2)} \right) \right) / 2 \right]^{1/2}$$

where

$$B_0 = 8\pi\omega\mu\eta\lambda_L^2 / \phi_0$$

In the two foregoing equations, $f$ is the fraction of free fluxons at an induced flux $B$, $\eta$ is the fluxon viscocity, $\lambda_L$ is the London penetration depth, $\omega$ is the rf frequency, $\mu$ is the permeability of the sample. As can be seen from above equations that the higher $\eta$ value decreases $R_s$, the dissipative parameter in the flux flow model [14] and make the variation of $R_s$ slower as a function of magnetic field. This might be the reason why we don't see any magnetic field dependent NRRA signals due to flux flow in $MgB_2$ at temperature below $T_c$. And due to the presence of less number of Josephson junctions, the intensity of NRRA signal (which is due to JJ decoupling) of the $MgB_2$ pellet is lesser than other high $T_c$ materials of same size.

Three phase reversals in $MgB_2$:

Figure 2 shows a clear picture of the magnetic field dependent NRRA signals of $MgB_2$ polycrystalline samples at some representative temperatures below $T_c$. The most interesting aspect of the NRRA signals in $MgB_2$ is the presence of three phase reversals, which is not seen in any other superconductors before. The phase of the temperature dependent NRRA signals goes through three separate and distinguished phase-reversals at 36, 27 and 20 K. We observe a single signal of the "correct phase" at temperatures down to 37 K below $T_c$, indicating that the actual absorption is a minimum at zero field and increases with an increase in the field. This conclusion regarding the phase was reached after noting that this signal is opposite in phase to $^1H$ (proton) NMR signal in a sample of glycerin. When the sample is cooled further down to 36 K, another narrower signal develops centered at zero field with its phase being opposite to the phase of the main signal. We designate this as the "first anomalous signal", is clearly visible at 30 K. When the sample is cooled further to 27 K, another narrower signal develops centered at zero field with its phase being same to the phase of the main signal. We designate this as the "second anomalous signal", is clearly visible at 24 K. When the sample is cooled further to 20 K, another narrower signal develops centered at zero field with its phase being opposite to the phase of the main signal. We designate this as the "third anomalous signal", is clearly visible at 16 K. Further decrease in temperature down to the lowest does not result in any major change in the line shape except for a decrease in the intensity of the signal in the temperature range 20 to 4 K. We note that these behaviors of the reverse phase signals are unlike that observed in ceramics [21], single crystal [22] and thin film [23] samples of HTSC in microwave frequencies, where a complex temperature dependent evolution of the line shape was observed.

The ceramic samples consist of superconducting grains of size of about a few microns (here it is less than 10 μm >> $\xi_0$, the coherence length) and, therefore, with well-defined magnitude of the superconducting order parameter. The phase of the order parameter remains random. These superconducting grains are now weakly coupled through normal intergranular regions, giving superconducting-normal-superconducting (S-N-S) junctions. The coupling energy is, however, distributed randomly over a certain range. Two neighboring grains can become phase locked if the intergranular coupling energy ($J_c\phi_0/2\pi$) exceeds the thermal energy, ($K_BT$), where $J_c$ is the critical current of the intergranular junction. Thus as we lower the temperature, more and more grains become phase locked and beyond a bond-percolation threshold the bulk superconductivity sets in.

This is the familiar bond-percolation model [2]. It is to be noted that the strength of the bond, that is, the coupling energy, is a sensitive function of magnetic field and temperature. The application of magnetic field and the variation of temperature alter the ac penetration depth of the sample, which in turn change the NRRA signals associated with it. This is the reason why the NRRA signals show the "correct phase" in the transition region (37 to 39 K). At other temperatures (below 37 K), frequent JJ decoupling is the dominating factor. The Josephson junction critical current decreases with increasing magnetic field, so the decoupling energy ($E_j$) decreases with increasing magnetic field. Due to the decreasing of $E_j$ with increasing magnetic field, NRRA signals at temperatures below 37 K show the "opposite phase" down to 30 K, which shows a decrease in the dissipation with increasing magnetic field. The NRRA signals at different temperatures from 30 to 4 K are attributed with the two band-gaps in MgB$_2$ [24, 25]. Due to the presence of two band-gaps ($\Delta_\sigma$ and $\Delta_\pi$) in MgB$_2$, Josephson junction critical current among the superconducting grains varies as $J_\sigma$ and $J_\pi$. The presence of two different Josephson junction critical current ($J_\sigma$ and $J_\pi$) makes two separate groups of weak links inside the polycrystalline sample. This two separate groups of weak links, which can have two different decoupling energy ($E_{j\sigma}$ and $E_{j\pi}$), lead to two more phase reversals with further decreasing in temperature below 30 K.

Effect of MgO doping:

Figure 4 shows a clear picture of the NRRA signals recorded from the MgO doped MgB$_2$ polycrystalline samples at different temperatures below $T_c$. The transition temperature of MgO mixed MgB$_2$ is 35 K (4 K less than pure MgB$_2$). The signals show very strong hysteresis, which increases with increasing temperature upto 24 K and disappears at $T_c$. The JJ signals, which present in MgB$_2$ granular sample disappears by the MgO doping.

The structure of MgB$_2$ can be easily tuned in two ways: chemical doping and external pressure. In chemical doping, by selecting different doping elements and doping concentration, the average B site atom size $<r_B>$ in the MgB$_2$ system is changed. Because of the mismatch between $<r_M>$ and Mg site ion, the local atomic structure of MgB$_2$ can be modified. MgO doping in MgB$_2$ results in highly anisotropic lattice contraction and a depression of $T_c$. It has been reported in some of the previous works that the presence of MgO causes the $J_c$ of the MgB$_2$ films to increase [26]. This has been explained in terms of increased pinning due to MgO insulating regions in the sample. We observe the same effect in the NRRA measurements. The absence of the JJ decoupling signals is due to the enhancement in the $J_c$ value. Due to the large enhancement in the Josephson junction critical currents of the Josephson junctions formed inside the samples, the rf current become unable to break those Josephson junctions and because of this the JJ decoupling signals are absent in MgO doped MgB$_2$. The strong hysteresis behavior in MgO doped MgB$_2$ sample is discussed next.

Hysteresis behaviors:
Both the samples, MgB$_2$ and the MgO doped MgB$_2$ show strong hysteresis in the NRRA signals. The hysteresis behavior of MgB$_2$ sample as a function of temperature is shown in Figure 5 and of MgO doped MgB$_2$ sample is shown in Figure 6. In the case of MgB$_2$, hysteresis increases with increasing temperature upto 38 K and after that it suddenly decreases. In the case of MgO doped $MgB_2$, hysteresis increases with increasing temperature upto 24 K and after that it slowly decreases. The hysteresis increases because the development of intergranular currents allows new pinning of fluxons. At still higher temperature fluxon pinning reaches a maximum when the intergranular currents are maximized and beyond that the intergranular currents and the hysteresis decreases. This is due to the dipinning of fluxons with thermal excitation. The stronger hysteresis in the case of MgO doped $MgB_2$ than $MgB_2$ is due to the presence of MgO, which act as the insulating region between superconducting grains where the chances of field being trapped increases.

Effect of sintering at $650^0$ C:

Figure 7 shows the NRRA signals of both the sintered and non-sintered $MgB_2$ pellets at 10 K. The magnitude of NRRA signal of the non-sintered $MgB_2$ pellet is nearly 6 times more than the sintered sample. Oxygen stoichiometry, grain size and intergranular and intragranular contacts are found to affect the NRRA signal. The annealing temperature of $650^0$ C was rather high to make a denser microstructure. The dense microstructure in the sintered $MgB_2$ sample reduces the number density of JJs present in the sample and also makes the existing JJs stronger. The granular pellet is more favorable for weak JJs. So in the granular pellet, the number density of JJ is expected to be more than the sintered pellet. Due to the occurrence of a large number of weak Josephson junctions in the granular $MgB_2$, JJ decoupling in the granular $MgB_2$ pellet is more frequent than the sintered pellet. The frequent JJ decoupling is the source of the NRRA signal in the polycrystalline MgB$_2$ samples; so the NRRA signal amplitude is more in the granular MgB$_2$ pellet than the sintered MgB$_2$ pellet.

Effect of grain size and pressure:

Figure 8 shows the details of the NRRA signals of granular MgB$_2$ made of grains of five different sizes; <10 μm (a), <15 μm (b), <20 μm (c), <25 μm (d), and <30 μm (e). The NRRA signals amplitude decreases with decreasing grain size. Figure 9 shows the detail of the NRRA signals of granular MgB$_2$ pellet made at three different pressures; 10 MPa, 20 MPa and 30 MPa. The NRRA signals amplitude decreases with increasing pressure. Increasing pressure or decreasing grain size, both have similar effects on the NRRA signals in granular samples. Increasing pressure or decreasing grain size decreases the number of weak links inside the pellets and makes some of the weak links stronger than before. In the first case, the NRRA signal amplitude decreases because of the reduction in the total number of Josephson junctions and in the second case the NRRA signal amplitude decreases because some of the weak links don't break with the same amplitude of applied rf current. It is reported that the $T_c$ of MgB$_2$ material changes at very high pressure but we did not observe any noticeable change in the $T_c$ with pressures upto 30 MPa.

Effect of rf power and modulation amplitude:

Figure 10 shows the NRRA amplitude of MgB$_2$ granular sample as a function of the rf power at two different field modulations (10 G and 100 G) at 10 K. The trends are quite different for these two field modulations, which show that the NRRA response is rather sensitive to rf power. The rf power dependence can be explained in terms of intergranular and intragranular currents. The total surface current generates a magnetic field that opposes the rf magnetic field and decreases the rf penetration depth. For small field modulation, only intragranular currents are active. Consequently, the maximum NRRA amplitude at which the critical current through the Josephson junction is reached occurs at low rf power. For large field modulation both intergranular and intragranular currents are active and the maximum NRRA amplitude is reached at much higher rf power.

Figure 11 shows the NRRA amplitude as a function of modulation amplitude at 6 and 12 K. The linear increase with modulation amplitude upto 20 G can be explained by the total surface current. The total surface current is a combination of intergranular and intragranular currents. At field modulation the intergranular currents are generated within individual grains. These intergranular currents increase on increasing the modulation amplitude upto 20 G give an increase in the NRRA signal amplitude. Increasing the modulation amplitude further above 20 G does not show linear behavior and saturates at higher modulation amplitudes.

Effect of passing dc bias current:

The NRRA signal patterns of $MgB_2$ sample are studied in the presence of dc bias currents. When a dc current passes through the superconducting sample, changes in the phase of the NRRA signal are observed. Figure 12 shows the NRRA signals of an $MgB_2$ sample of dimensions $15 \times 5 \times 4$ mm$^3$ with and without bias current at 10 K. The rf current passes through the sample has ability to break some of the weak Josephson junctions which have critical current less than the rf current. But in the presence of a dc current, the same rf current become able to break more number of the weak Josephson junctions and also able to break some of the strong Josephson junctions inside the sample. So passing dc current has the same effect as decreasing the Josephson junction coupling energy. We have already discussed that the occurrence of three phase reversals in $MgB_2$ is connected to the Josephson junctions decoupling energy and two band-gaps in $MgB_2$. Similarly passing dc current plays a major role in changing the Josephson junctions decoupling energy inside the samples, which can change the phase of the NRRA signal. Figure 13 shows the bias current dependence of the $dP/dH$ at -25 Gauss and 10 K. Here it shows how nicely the phase changes with increasing the current. The phase starts changing at dc current of 70 mA. The phase changes further with increasing current, which is not shown in the figure.

Frequency effect:

It is observed that in the NRRA experiment, the resonant frequency of the oscillator changes with the variation of temperature and the magnetic field. Figure 14 depict the isothermal field dependence of the frequency

$$f(H,T) = \frac{1}{2\pi [L(H,T)C]^{1/2}}$$

of the tank circuit with the sample inserted in the coil. Here $L$ is the effective inductance and $C$ is the capacitance in the tank circuit, The field dependence of $f(H)$ essentially arises due to the rf penetration (i.e., an increase in $L$) into the sample as $H$ is increased. At all the temperatures, the frequency remains maximized at zero field and decreases with increasing the magnetic field. Increase the field from $H=0$ onwards leads to the weakening of the screening rf currents and therefore the effective volume penetrated by the rf increases. The increase in field also begins to suppress the weak Josephson junction coupling between the grains in the sample. Consequently, the ability of the material to sustain a strong rf screening current through the sample and the decrease in frequency is observed. Increasing the field beyond has no further effect on the rf penetration and hence it remains independent of the field leading to the saturation of the frequency. Figure 15 shows the appearances of hysteretic behavior in the frequency variation at zero field. The hysteretic behavior in the resonant frequency is the consequence of the trapped flux.

Effect of sample size:

The NRRA is a very sensitive technique for the determination of $T_c$ in superconducting samples. In our rf experiment we found that the transition temperature $T_c$ of granular $MgB_2$ samples measured using the NRRA technique decreases with decreasing sample size, whereas the $T_c$ measured using ac susceptibility and temperature dependent resistivity of same samples is independent of the sample size. This phenomenon has not been observed in either high temperature or low temperature superconductors so far. It is very important to clarify this problem. On the other hand, we need to understand the underlying mechanism governing this dependence in order to see whether we can correlate all the three parameters, which vary with the sample size in $MgB_2$, the critical current density, the activation energy, and the $T_c$ determined from our NRRA techniques.

The NRRA signal tells the overall response of sample in the presence of rf. The $MgB_2$ samples used in this study are all of rectangular shapes. In order to eliminate any geometrical effect on $T_c$, the pellet was cut into a series of samples with constant size ratio a: b: c; here 'a' is the length, 'b' is the breadth and 'c' is the height of the rectangular samples. Seven polished rectangular samples were used in this study with dimensions of a × b × c mm$^3$ = 18 × 9 × 9, 15 × 7 × 7, 12 × 6 × 6, 10 × 5 × 5, 7 × 3.5 × 3.5, 6 × 3 × 3, 5 × 2.5 × 2.5 mm$^3$. The dependence of $T_c$ on sample sizes is shown in Figure 16. It is observed that the transition temperature $T_c$ measured using the NRRA technique decreases with decreasing sample size (From 18 mm to 5 mm size), whereas the $T_c$ measured using ac susceptibility and the temperature dependent resistivity is independent of the sample size. The above results were not found in other superconductors like YBCO, LSCO and BSCCO with varying sample size (From 18 mm to 5 mm).

In MgB$_2$ sample we concluded that due to strong pinning we don't see any NRRA signal due to flux motion. The entire signals are due to the JJ decoupling. Above the temperature $T_d$, where the intergranular coupling energy ($J_c\phi_0/2\pi$) exceeds the thermal energy ($K_BT$), all the JJs present in the sample are decoupled and the loss due to JJ decoupling is saturated. Above the temperature $T_d$ there wont be any JJ left in the samples to show magnetic field dependent loss due to JJ decoupling. In MgB$_2$ sample, the disappearance of the central $dP/dH$ signal for bigger samples at higher temperature indicates that the temperature $T_d$ is higher for bigger samples as compared to the smaller samples. This also indicates that the coupling energy, which is described in the above BP model, is higher for bigger samples as compared to the smaller samples. Due to the above reasons, the $T_c$ measured by NRRA experiment depends on sample sizes. It is very difficult to give a quantitative picture for the variation of JJ decoupling energy with sample size. Qualitatively we can say that when the sample size becomes bigger then the possibilities of bigger and stronger JJs among the grain boundary weak links increase. Bigger JJs have higher critical current than smaller JJs and hence higher decoupling energy.

In case of superconductors like YBCO, LSCO and BSCCO, we don't see any unusual dependence of $T_c$ on sample sizes of similar range due to two reasons; 1: JJ decoupling energy is larger than $MgB_2$ and the magnitude never comes below the thermal energy ($K_BT$). 2: Low activation energy.

Calculation of coupling energy:

From the above discussion it is clear that the temperature at which the NRRA signal disappears in $MgB_2$ is the temperature where intergranular coupling energy becomes equal to the thermal energy. So in $MgB_2$ intergranular coupling energy varies from 25 to 39 $K_BT$ when we vary the sample (pellets are made of grains less than 10 μm in size, and pressurized at 30 MPa) size from $5 \times 2.5 \times 2.5$ to $18 \times 9 \times 9$ mm$^3$. This coupling energy is very small compare to other high $T_c$ superconductors. Due to this low intergranular coupling energy, $MgB_2$ does not exhibit weak-link electromagnetic behavior at grain boundaries properly and because of this in the literature; $MgB_2$ was mentioned as the absence of weak links [11]. The JJ decoupling energy is completely associated with the grain size, the distance between grains and the contact area between grains. So the rf responses will be more interesting in the case of MgB$_2$ made of nano size particles, which experiments are beyond our experimental facilities.

**Conclusions:**

To summarize, a close correlation between the NRRA signals and the microstructures of the superconducting MgB$_2$ samples is observed. The anomalous three phase reversals in the granular MgB$_2$ are attributed to due to the Josephson junction decoupling and the presence of double band-gaps. Sintering at $650^0$ C for 10 hours, increasing pressure or decreasing grain size reduces the NRRA signal amplitude in the granular MgB$_2$ sample. This is due to the reduction in the total number of available Josephson junctions and increasing in the $J_c$ values in some of these JJs. Increasing the rf amplitude above a certain critical value can make the amplitude of NRRA signal saturates and further increasing in the rf power results in decreasing in the amplitude of NRRA signal in the superconducting state. The NRRA signal shape is sensitive to the pinning and depinning of magnetic fluxons and Josephson junctions decoupling. The magnitude of the NRRA signal depends on the sum of intergranular and intragranular currents through Josephson junctions and the rf penetration depth. Both the factors are dependent on the density of the material. Conditions favoring hysteresis are, MgO doping, low field modulation, low rf power and high temperature. The $T_c$ measured using the NRRA technique reduces with reducing sample size above collective pinning length. Collective pinning length $L_c$ and the activation energy $U_c$ in the direction parallel to the field are found to be higher than other superconductors like YBCO, LSCO and BSCCO. The intergranular coupling energy varies with sample sizes and is very less compared to other high $T_c$ cuprate superconductors. The NRRA response in MgB$_2$ sample in the superconducting state is completely due to intragranular and intergranular JJ decoupling and is not due to the vortex movement. MgB$_2$ can be a better candidate for making rf devices than other HTSC like YBCO, LSCO and BSCCO due to lower rf loss.

**Acknowledgements:**

This work is supported by the Department of Science and Technology, University Grants Commission and the Council of Scientific and Industrial Research, Government of India.

**Figure Captions**:

1. Schematic of the low temperature rf probe showing the sample region.

2. Typical NRRA signals recorded from a sample of $MgB_2$ (grain size < 10 μm and pressurized at 30 MPa) during heating at few representative temperatures for both the forward and reverse scans of the magnetic field. The hysteresis between the forward and reverse scans of the magnetic field shows near field independence, thus forming practically rectangular hysteresis loop. The $T_c$ (~ 39 K) is the temperature where the NRRA signal disappears. The arrows show the forward and reverse directions of magnetic field sweep.

3. Typical NRRA signals of an YBCO granular pellet at different temperatures below $T_c$. The magnetic field is scanned in the forward directions from -150 to 150 Gauss. Just few degree below $T_c$ there is a clear indication of NRRA signals due to both Josephson junction decoupling and flux motion. The arrows showing the peaks in the NRRA signals due to the flux motions.

4. Typical NRRA signals recorded from a sample of MgO doped $MgB_2$ during heating at a few representative temperatures for the both forward and reverse scans of the magnetic field. The hysteresis between the forward and reverse scans of the magnetic field shows near field independence, thus forming practically rectangular hysteresis loop. The $T_c$ (~ 35 K) is the temperature where the NRRA signal disappears. The arrows show the forward and reverse directions of magnetic field sweep.

5. Temperature dependence hysteresis of $MgB_2$. The area under the forward and reverse sweep of the magnetic field is a measure of hysteresis. The magnetic field is scanned in both the forward and reverse directions from -150 to 150 Gauss.

6. Temperature dependence hysteresis of MgO doped $MgB_2$. The area under the forward and reverse sweep of the magnetic field is a measure of hysteresis. The magnetic field is scanned in both the forward and reverse directions from -150 to 150 Gauss.

7. NRRA signals of both the granular and sintered $MgB_2$ pellet at 10 K. It is important to note that the NRRA signal of granular sample reduces after sintering at $650^0$ C for 10 h. The magnetic field is scanned in the forward directions from -150 to 150 Gauss.

8. NRRA signals of $MgB_2$ granular samples (pressurized at 10 MPa) for five different grain sizes at 10 K: (a) less than 10 μm; (b) less than 15 μm; (c) less than 20 μm (d) less than 25 μm; (e) less than 30 μm. The magnetic field is scanned in the forward directions from -150 to 150 Gauss

9. NRRA signals of $MgB_2$ granular samples (grain sizes less than 30 μm) for three different pressures (30 MPa, 20 MPa, and 10 MPa) at 10 K. The magnetic field is scanned in the forward directions from -150 to 150 Gauss

10. RF power dependence of NRRA signal amplitudes of $MgB_2$ granular sample for two different field modulations at 10 K.

11. Magnetic field modulation dependence of NRRA signal amplitude of $MgB_2$ granular sample at two different temperatures. The modulation frequency was set at 100 Hz and the modulation amplitude was varied from 1 to 20 Gauss. The rf power was set at 250 mW.

12. Effect of passing dc current on the NRRA signal of $MgB_2$ granular sample at 10 K. The phase of the sample changes at 75 mA dc current.

13. The current dependence of the dP/dH at -20 Gauss magnetic field and 10 K. The phase starts changing at dc current of 70 mA. Increasing the current above 100 mA can change the phase of the signal further.

14. Magnetic field dependence of resonant frequency of the Robinson oscillator at different temperatures below $T_c$. The resonant frequency decreases with increasing magnetic field or temperature. The magnetic field is varied upto 1 T.

15. Magnetic field dependence of resonant frequency of the Robinson oscillator at two temperatures (6 K and 25 K). The resonant frequency decreases with increasing magnetic field or temperature. The magnetic field is varied upto 3000 G. There is a clear indication of hysteresis near to the zero field at both the temperatures.

16. Transition temperature $T_c$ measured using NRRA technique versus the sample size. The sample size corresponds to the length of the cylindrical $MgB_2$ samples. It is important to note that the $T_c$ determined from the NRRA experiment decreases with decreasing sample size whereas the $T_c$ measured using ac susceptibility and temperature dependent resistivity is independent of the sample size.

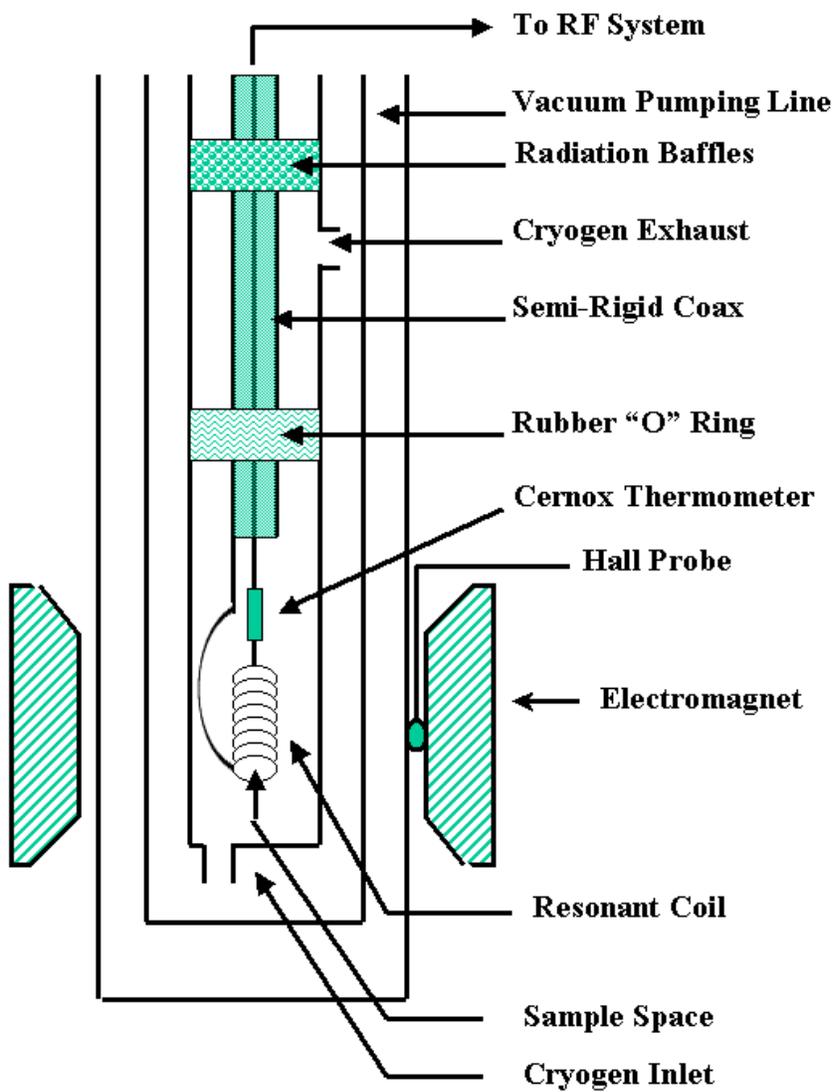

**FIG. 1.**

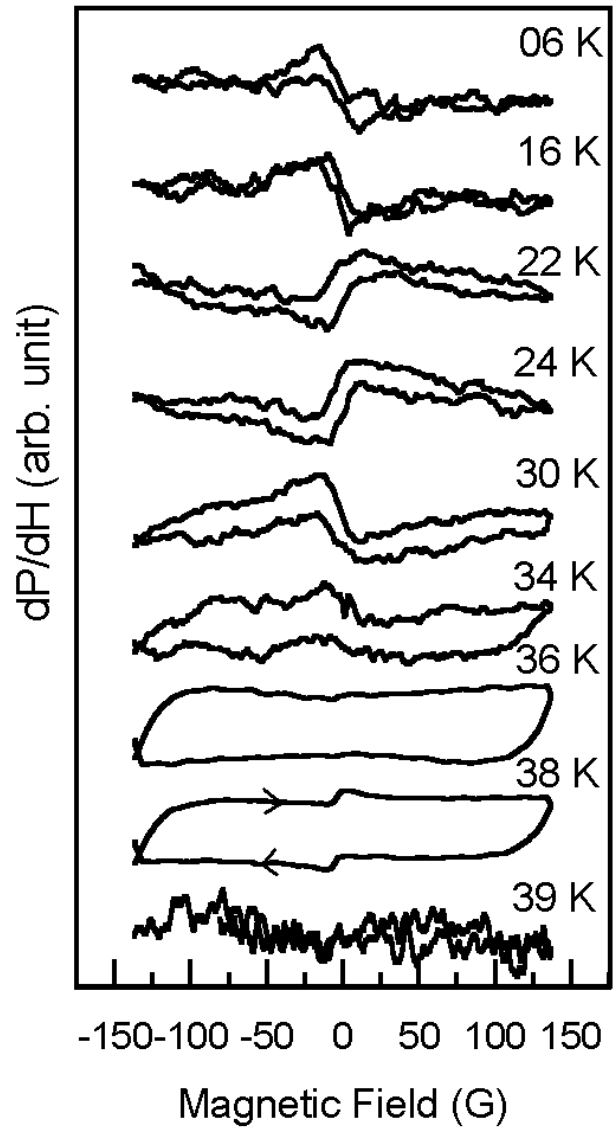

**FIG. 2.**

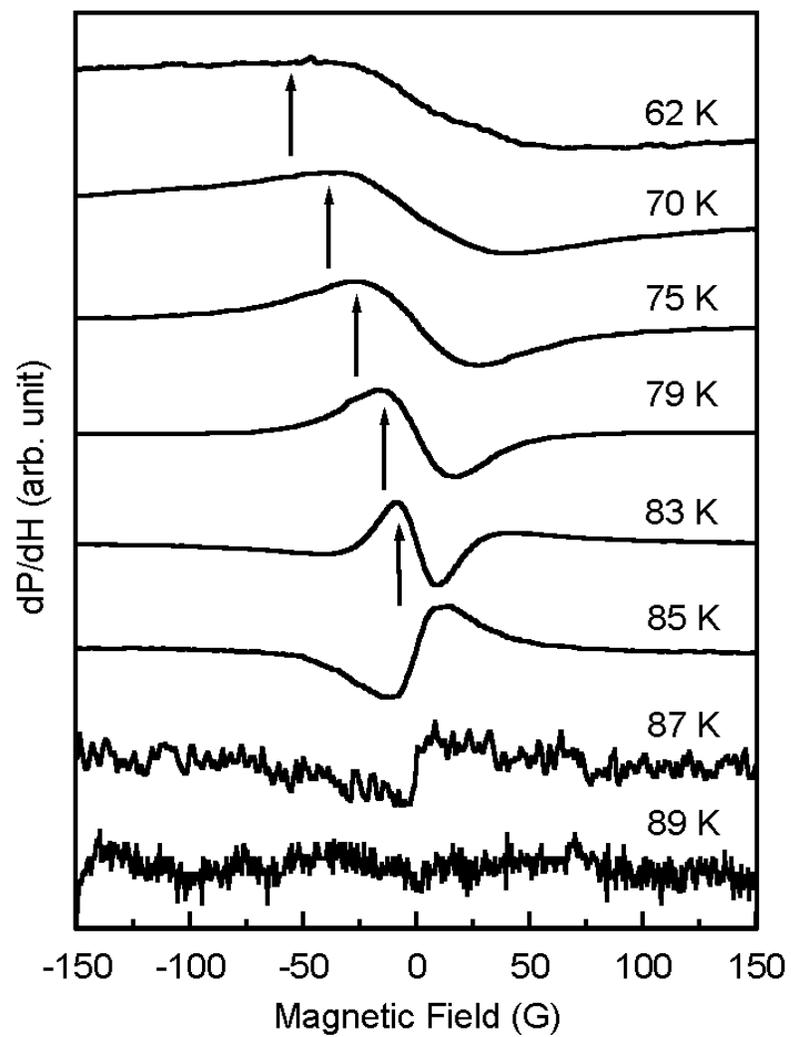

**FIG. 3.**

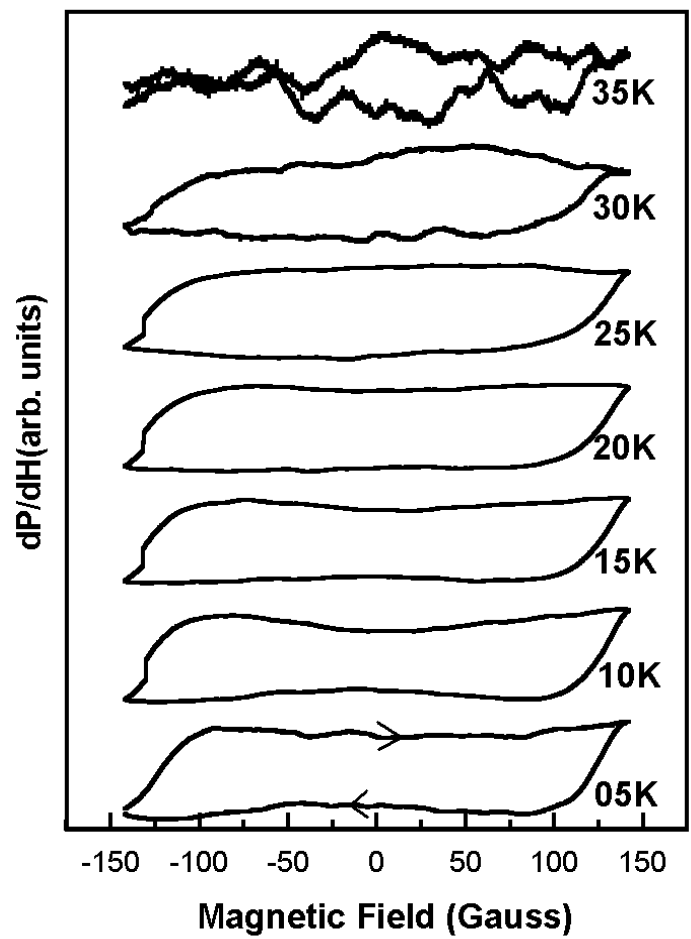

**FIG. 4.**

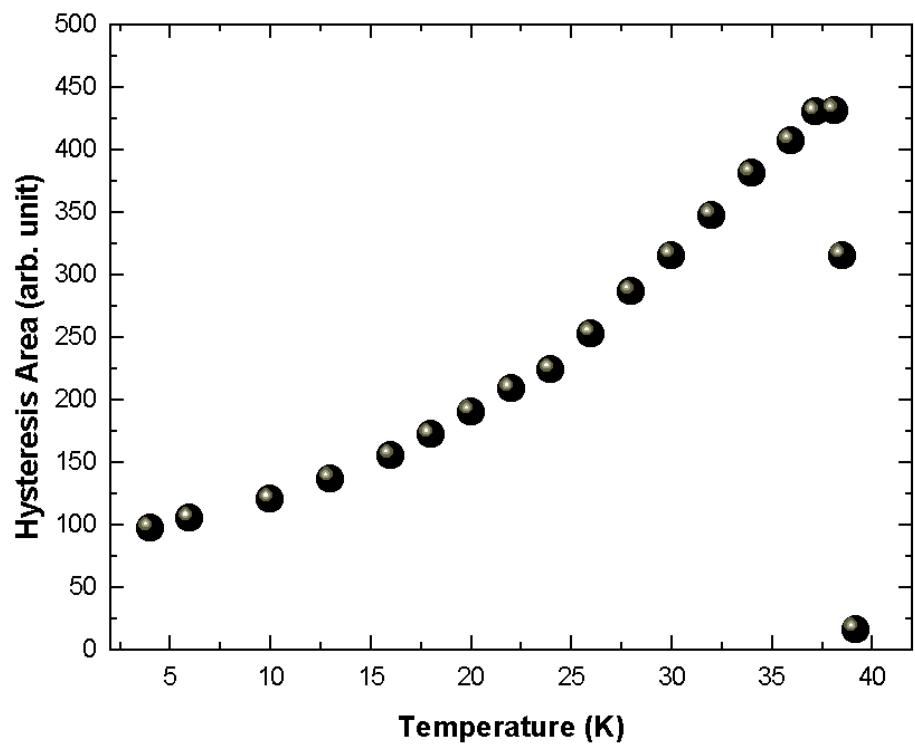

**FIG. 5.**

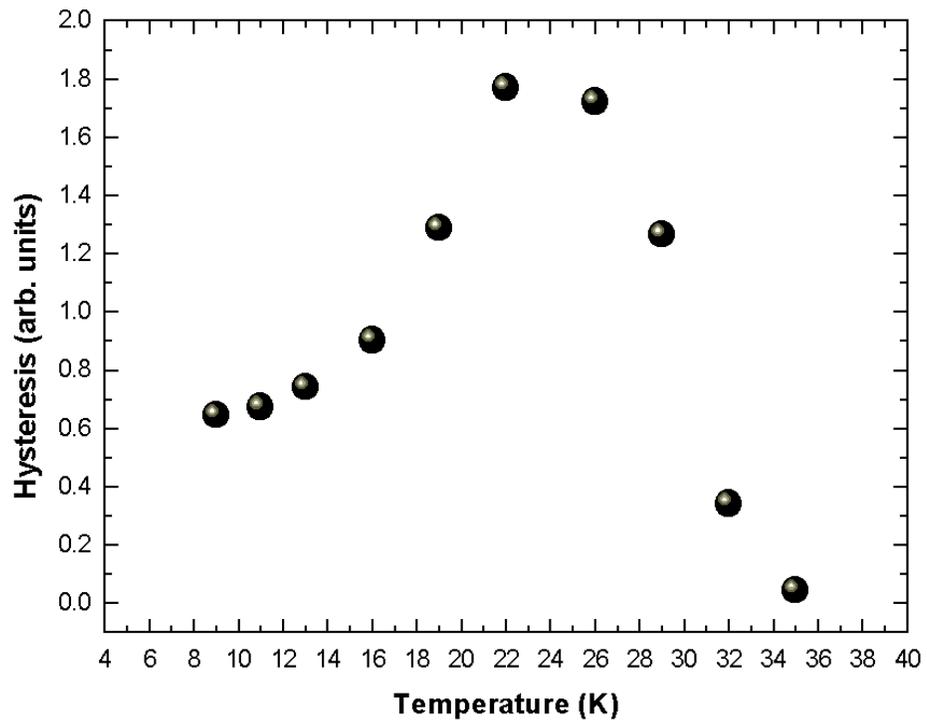

**FIG. 6.**

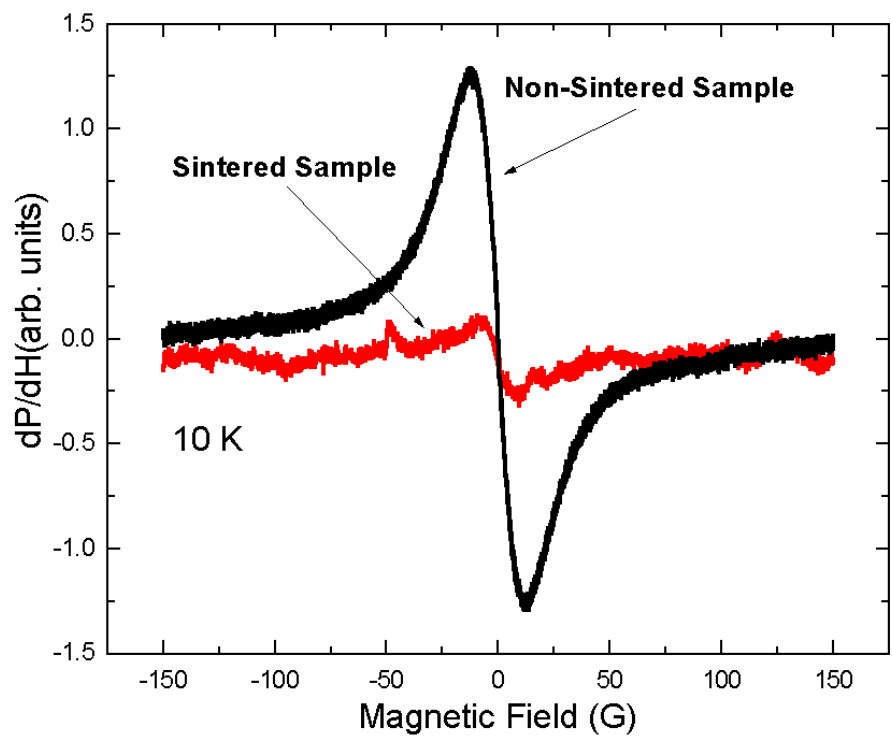

**FIG. 7.**

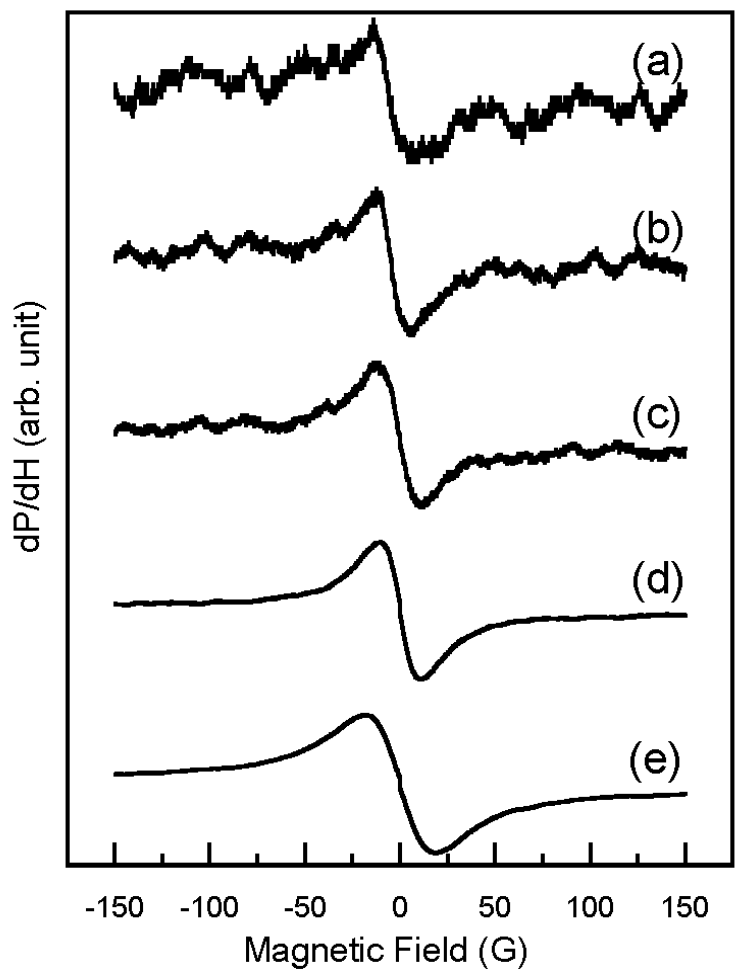

**FIG. 8.**

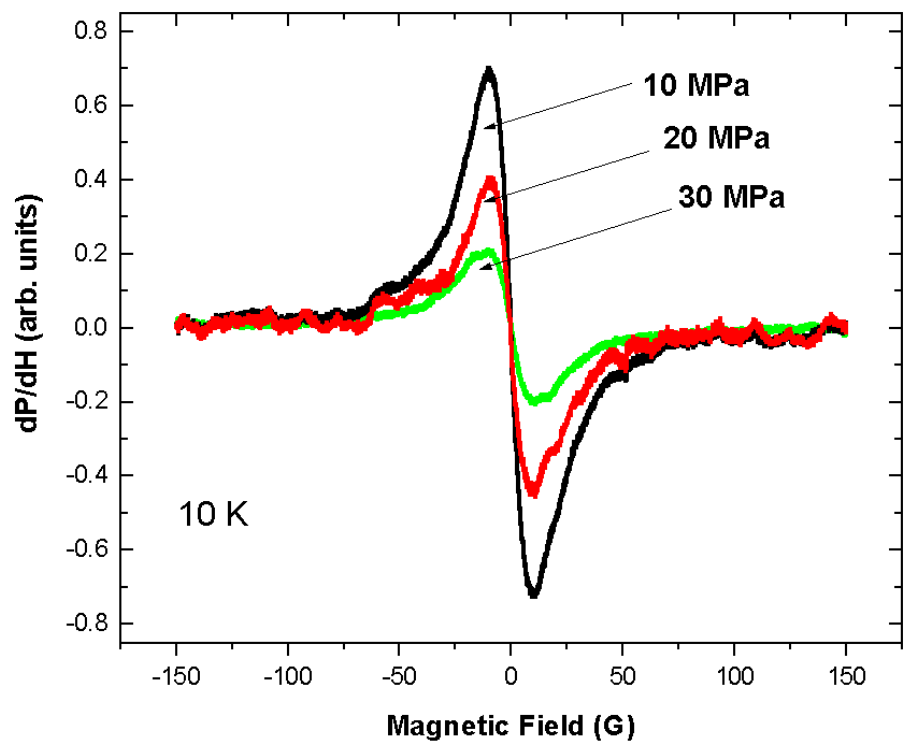

**FIG. 9.**

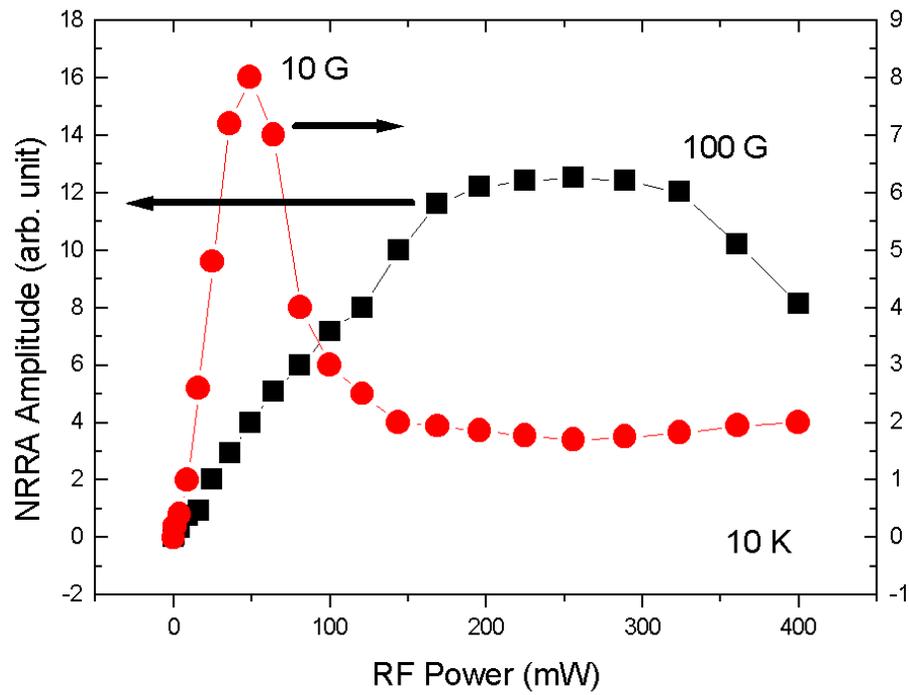

**FIG. 10.**

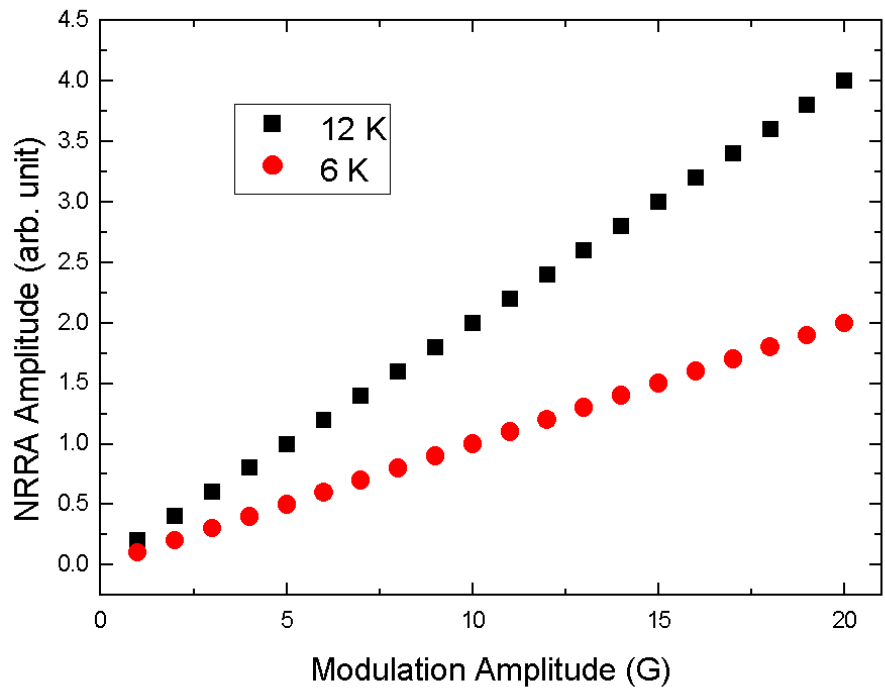

**FIG. 11.**

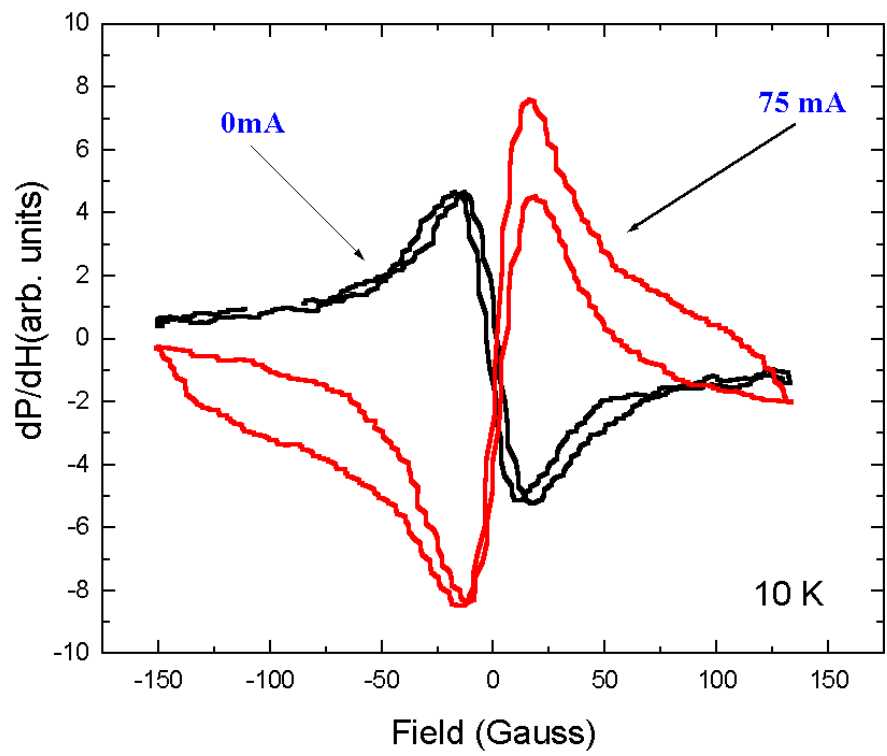

**FIG. 12.**

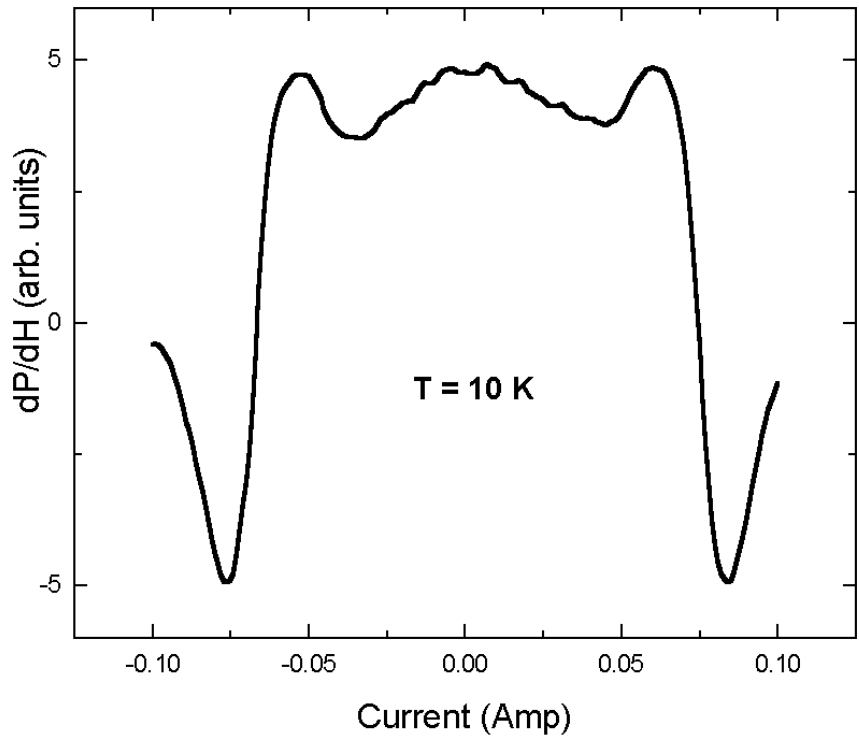

**FIG. 13.**

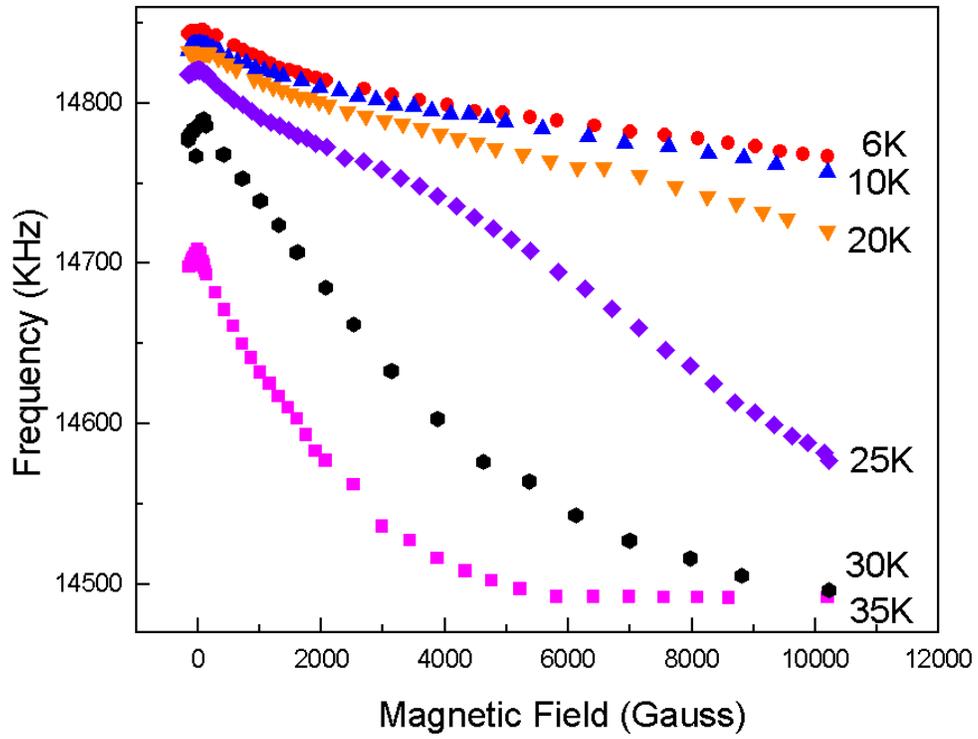

**FIG. 14.**

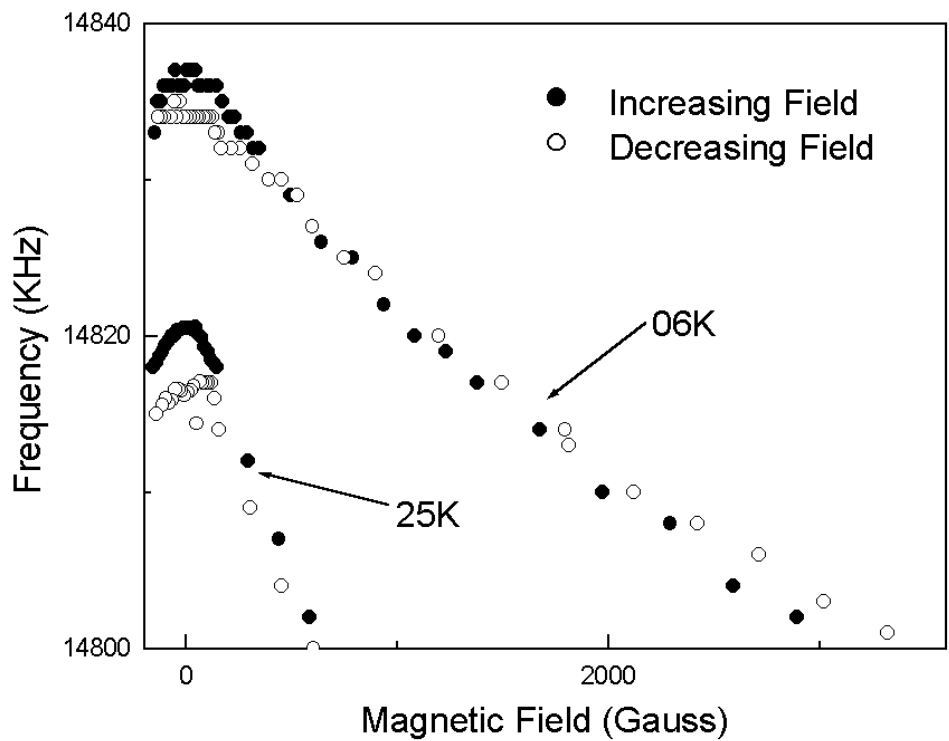

**FIG. 15.**

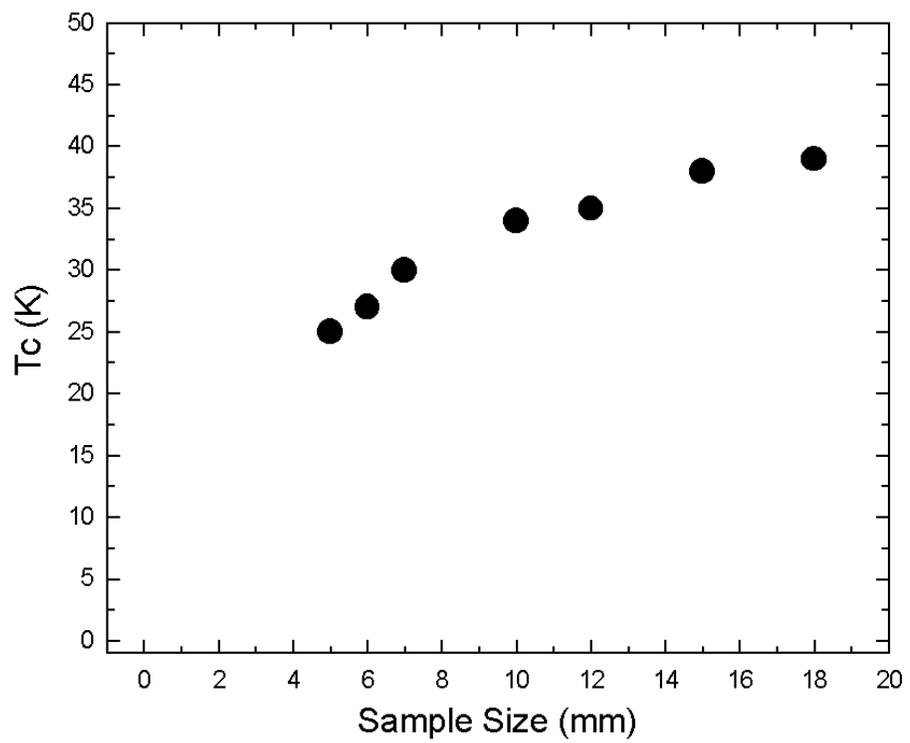

**FIG. 16.**